\documentclass[aps,prl,twocolumn,amsmath]{revtex4-1} %

\usepackage{ulem}

\usepackage{color,graphicx}
\usepackage{bm}%

\usepackage{amsmath}
\usepackage{amsfonts}
\usepackage{amssymb}

\usepackage{subfigure}
\usepackage[latin1]{inputenc}
\newcommand{\ket}[1]{\ensuremath{\left| #1 \right\rangle}}

\renewcommand{\-}{\,-\,}

\newcommand{\br}{\mathbf{r}}
\newcommand{\bk}{\mathbf{k}}

\let\oldmarginpar\marginpar
\renewcommand\marginpar[1]{\-\oldmarginpar[\raggedleft\tiny #1]%
{\raggedright\tiny #1}}

\begin{document}

\graphicspath{{./}}

\title{Adiabatic continuation of Fractional Chern Insulators to Fractional Quantum Hall States}

\author{Thomas Scaffidi}
\affiliation{Ecole Normale Sup\'erieure, 24 rue Lhomond, 75005 Paris, France}
\affiliation{Rudolf Peierls Centre for Theoretical Physics, Oxford University, Oxford OX1 3NP, United Kingdom}

\author{Gunnar M\"{o}ller}
\affiliation{TCM Group, Cavendish Laboratory, J.J.~Thomson Avenue, Cambridge CB3 0HE, UK}

\date{\today}

\begin{abstract}
We show how the phases of interacting particles in topological flat bands, known as fractional Chern insulators, can be adiabatically connected to incompressible fractional quantum Hall liquids in the lowest Landau-level of an externally applied magnetic field. Unlike previous evidence suggesting the similarity of these systems, our approach enables a formal proof of the equality of their topological orders, and furthermore this proof robustly extends to the thermodynamic limit. We achieve this result using the hybrid Wannier orbital basis proposed by Qi [Phys.~Rev.~Lett.~{\bf 107}, 126803 (2011)] in order to construct interpolation Hamiltonians that provide continuous deformations between the two models. We illustrate the validity of our approach for the groundstate of bosons in the half filled Chern band of the Haldane model, showing that it is adiabatically connected to the $\nu=1/2$ Laughlin state of bosons in the continuum fractional quantum Hall problem.
\end{abstract}

\maketitle

Owing to the recent discovery of topological insulators \cite{Hasan:2010p826}, there is now hope to realize materials which manifest Haldane's vision of a quantum Hall effect without external magnetic fields \cite{Haldane:1988p634,Yu:2010p3287}.
Several proposals have extended this concept to fractional quantum Hall (FQH) liquids that could be realized in topologically non-trivial  bands which are also flat \cite{Tang:2011p1780,Sun:2011p1781,Neupert:2011p1803,Roy:2011p1779,[{See also, }][]Levin:2009p2889}. 
A similar mechanism was proposed to simulate the effect of strong magnetic fields in cold atomic gases \cite{Lin:2009p33,Cooper:2011p1498}. 
Flat bands with non-zero Chern number \cite{Thouless:1982p80,Niu:1985p530} provide an avenue to realize strongly correlated states at high temperatures \cite{Tang:2011p1780}, raising prospects for stabilizing exotic non-abelian phases required to build topological quantum computers \cite{Kitaev:2003p611}.

Numerical works seeking evidence for incompressible quantum liquids in topological flat bands have focused on spin polarized models breaking time-reversal symmetry \cite{Sheng:2011p2299,Wang:2011p2572,Regnault:2011p2571}, which were baptized as fractional Chern insulators (FCI) \cite{Regnault:2011p2571}. Signatures for the topological nature of their ground states include their spectral flow and groundstate degeneracies \cite{Sheng:2011p2299,Wang:2011p2572} and the analysis of the entanglement spectra. The latter reveal a counting of excitations matching that of FQH states at the corresponding band filling, e.g.~the Laughlin and Moore-Read \cite{Regnault:2011p2571} or Jain states \cite{Liu:2012p3286}. One drawback is that these data can be acquired only for finite size systems.

Current insights in the analytic theory of fractional Chern insulators rely on the analysis of the projected density operator algebra \cite{Parameswaran:2012p3290,Goerbig:2012p2703} or of emergent symmetries in the exact many-body spectrum \cite{Bernevig:2012p2858}.
There are several proposals for constructing FCI wave functions \cite{Qi:2011p1804, Lu:2012p3291, Mcgreevy:2012p3288, Murthy:2011p2119,Vaezi:2011p3336,Wu:2012p3204,Murthy:2012p3282}, and some overlaps were calculated for fermionic systems \cite{Wu:2012p3204}. However, the understanding of the many-body ground states of FCIs cannot yet pride itself with an achievement similar to the celebrated accuracy of analytical wave functions for FQH states \cite{Laughlin:1983p301,Jain:1989p294,Moore:1991p165}.

In this manuscript, we provide a formal proof that FCIs are in the same universality class as FQH states. We base our argument on Qi's proposal \cite{Qi:2011p1804} of a mapping between FQH and FCI wavefunctions. By representing both FCI and FQH Hamiltonians in a Hilbert space with the same structure, we are able to study a class of superposition Hamiltonians that extrapolate smoothly between these systems. Taking advantage of this construction, we demonstrate that the many-body ground states of bosons in a half filled lowest Landau level and the topological flat band of the Haldane model are adiabatically connected, proving formally that these phases have the same type of topological order in the thermodynamic limit. Our strategy can be employed generally to identify incompressible quantum liquids in Chern bands, including bands with Chern number $|C|>1$.

Let us briefly comment on the case of lattice FQH states \cite{Sorensen:2005p58,Palmer:2006p63,Moller:2009p184,Hormozi:2012p3156}, which are simultaneously FCIs in the Chern bands of the Hofstadter butterfly \cite{Hofstadter:1976p69} given that flux can be gauged away. Nonetheless, these systems can be taken to the limit of the continuum fractional quantum Hall effect (FQHE) \cite{Hafezi:2007p67}. Hofstadter bands can be realized in cold atomic gases \cite{Dalibard:2011p2863}, which may also provide the most promising avenue for realising FCIs since topological flat bands require fine-tuned parameters that are efficiently controlled in these systems \cite{Goldman:2010p2890,Campbell:2011p2392,Cooper:2011p1498,Dalibard:2011p2863}.

We first establish our notations for the description of fractional Chern insulators. We consider finite two-dimensional lattices of $N_\text{cell}=L_1 \times L_2$ unit cells, spanned by lattice vectors $\mathbf{v}_i$ forming an opening angle $\gamma$, and we choose $\mathbf{v}_1 = \sin(\gamma) \mathbf{e}_x+\cos(\gamma)\mathbf{e}_y$ and $\mathbf{v}_2 = \mathbf{e}_y$. Lattice sites are located on $n_b$ sublattices $\alpha$ within the unit cell. %
In a finite system with periodic boundary conditions $\Psi(\br+L_i \mathbf{v}_i) =  e^{\mathrm{i}\phi_i}\Psi(\br)$, the reciprocal lattice consists of  discrete points $\mathbf{k}= \sum_i (q_i + \frac{\phi_i}{2\pi}) \mathbf{G}_i$, where $\mathbf{G}_1 = 2\pi \mathbf{e}_x/ L_1 \sin(\gamma)$ and $\mathbf{G}_2 = 2\pi [-\cot(\gamma)\mathbf{e}_x+\mathbf{e}_y]/L_2$ and we consider a rhomboid fundamental region with $q_i=0,...,L_i-1$.

For the moment, we consider the Hamiltonian of the infinite system in its momentum space representation $\mathcal{H}=\sum_{\mathbf{k}} \hat c^{\dagger}_{\mathbf{k},\alpha} h_{\alpha\beta}(\mathbf{k}) \hat c_{\mathbf{k},\beta}$, %
where $h(\bk)$ yields a topologically non-trivial flat band.
Let the Bloch functions $u$ and eigenenergies $\epsilon$ be determined by the corresponding eigenvalue equation $h_{\alpha\beta}(\mathbf{k}) u^n_\beta = \epsilon_n(\bk) u^n_\alpha$, introducing the band index $n$, and with the normalization $\sum_\alpha |u_\alpha^n(\mathbf{k})|^2 = 1$. In the following, we will denote the eigenstates as $| n, \bk \rangle = \sum_\alpha u^n_\alpha(k) \hat c^\dagger_{\bk,\alpha} | \text{vac.} \rangle$. 
The ensuing Berry connection $\boldsymbol{\mathcal{A}}(n,\bk) %
= -\mathrm{i} \sum_\alpha u^{n*}_\alpha(k)  \nabla_\bk u^n_\alpha(k)$ is a gauge dependent quantity. Physical observables such as the Berry curvature $\mathcal{B}(k) = \nabla_\bk \wedge \boldsymbol{\mathcal{A}}(k)$ and the resulting Chern index $C=\frac1{2\pi}\int_{BZ} d^2\mathbf{k} \,\mathcal{B}(\mathbf{k})$ are gauge invariant.

We now review Qi's proposal for mapping FCIs onto FQH states by a construction of Wannier states within a topological flat band \cite{Qi:2011p1804}. In his approach, we formulate hybrid Wannier wavefunctions $|W(x,k_y)\rangle$ that are localized only along the $x$-axis, while retaining translational invariance with a well defined momentum projection $k_y$ onto the $y$-direction \cite{[{See also, }][] Marzari:1997p2715,*Yu:2011p2711,*Soluyanov:2011p2859,*Soluyanov:2012p3292,*Huang:2012p2712}. We can think of these states as the simultaneous eigenstates of momentum $\hat P_y$ and the band-projected position operator $\hat X^{cg} = \lim_{q_x\to 0}\frac{1}{\mathrm{i}}\frac{\partial}{\partial q_x} \bar \rho_{q_x}$ \cite{Parameswaran:2012p3290}, satisfying
\begin{equation}
\label{eq:Position}
\hat{X}^{cg} \ket{W(x,k_y)} = [x - \theta(k_y)/2\pi] \ket{W(x,k_y)}.
\end{equation}
We adopt the explicit construction of the Wannier states in terms of the momentum eigenstates of $h(\bk)$ presented in Refs.~\cite{Qi:2011p1804,Barkeshli:2011p2526}, given by
$\ket{W(x,k_y)} \equiv \sum_{k_x=0}^{2\pi} f_{k_x}^{x,k_y} \ket{n=0,(k_x,k_y)} $, for $x=0,\ldots,L_1-1$, with
\begin{align}
\label{eq:defWannier} 
 f_{k_x}^{x,k_y} = \frac{\chi(k_y)}{\sqrt{L_x}} e^{-\mathrm{i}\int_0^{k_x} \mathcal{A}_x(p_x,k_y)dp_x-\mathrm{i} k_x\left(x-\frac{\theta(k_y)}{2\pi}\right)}.
\end{align}
This expression is related to a simple Fourier transform of the momentum eigenstates by additionally taking into account parallel transport of the phase along $k_x$ according to the Berry connection $\mathcal{A}_x$. The polarization $\theta(k_y)=\int_0^{2\pi} \mathcal{A}_x(p_x,k_y)dp_x$ is required to ensure periodicity of the state in $k_x$, enforcing $f^{x,k_y}_{k_x}=f^{x,k_y}_{k_x+2\pi}$. 
The relative phases $\chi(k_y)$ of the Wannier states represent a gauge freedom of the theory, while the relative phase of Bloch functions at the same $k_y$ is absorbed by the Berry connection in (\ref{eq:defWannier}). We take the particular choice
 $\chi(k_y)=  \exp[-\mathrm{i}\int_0^{k_y} \mathcal{A}_y(0,p_y)dp_y+\mathrm{i} \frac{k_y}{2\pi} \int_0^{2\pi} \mathcal{A}_y(0,p_y)dp_y ]$, as suggested in \cite{Barkeshli:2011p2526}.

 \begin{figure}
\begin{center}
\includegraphics[width=0.95\columnwidth]{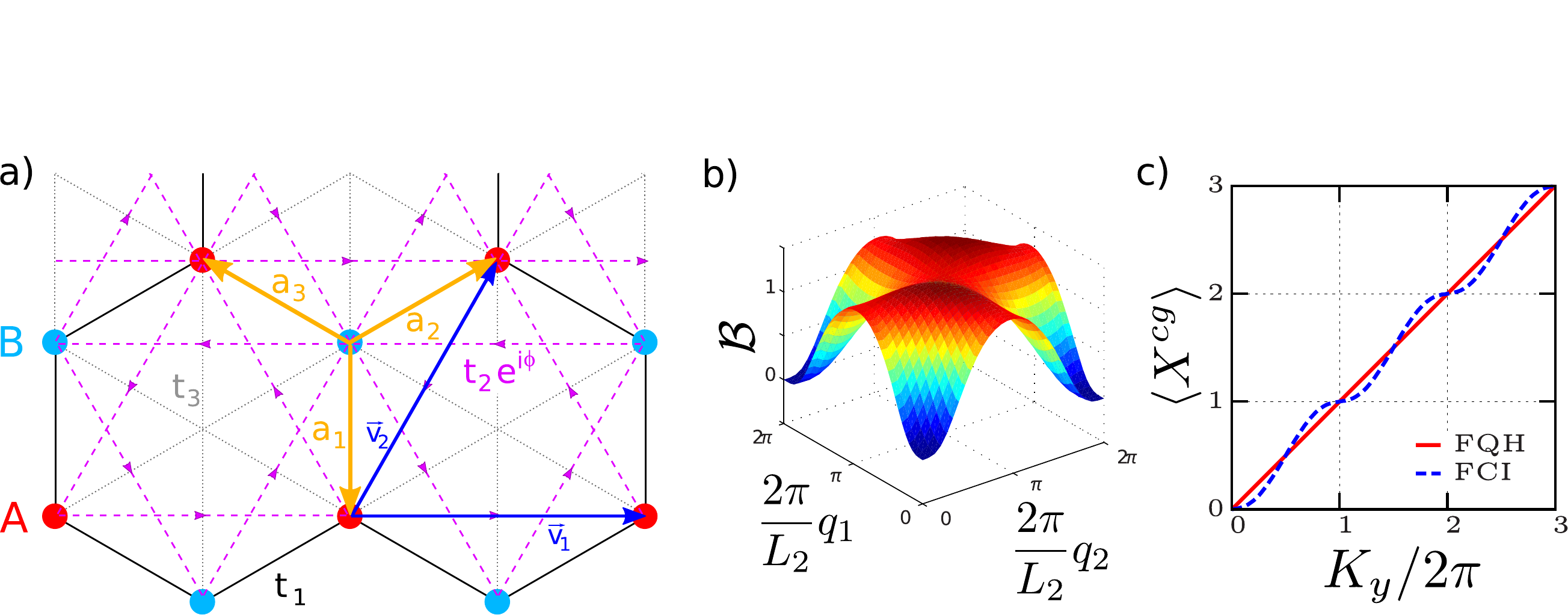}\\
\caption{\label{fig:HaldaneModel}
(a) Geometry and hopping terms in the Haldane-Model: the fundamental unit cell has two inequivalent sites $A$ and $B$. Second nearest neighbor interactions are complex, with arrows indicating the direction of a positive hopping phase $\phi_{\br\br'}$. (b) Berry curvature for the Haldane model. (c) Corresponding expectation value of the position operator $\langle \hat X^{cg}\rangle$ for Wannier states according to Eq.~(\ref{eq:Position}).%
}
\end{center}
\end{figure}

In finite-size systems, we adapt the construction (\ref{eq:defWannier}) straightforwardly. Given the Bloch functions $u_{\beta}^{n}(\mathbf{k})$ on reciprocal lattice points in the fundamental region, we choose a gauge that is consistent with the periodicity of momentum space, i.e. $u_{\beta}^{n}(\mathbf{k}+L_i\mathbf{G_i})=u_{\beta}^{n}(\mathbf{k})$.
A discretized version of the Berry connection of band $n$ is then computed as %
$A_x^n(q_1,q_2) = \Im \log \left[ u_{\alpha}^{n*}(q_1,q_2) u_{\alpha}^n(q_1+1,q_2)\right]$.
The integral of the Berry connection %
is discretized as 
$\int_0^{k_x} \mathcal{A}_x(p_x,k_y)dp_x \rightarrow \sum_{\tilde q_1=0}^{q_1(k_x)} A_x^n(\tilde q_1,q_2)$, 
and mutatis mutandis for $A_y^n(q_1,q_2)$. This resolution of (\ref{eq:defWannier}) yields a unitary transformation of the original single particle basis.
 As angles, the values of $A_x^n(q_1,q_2)$ are defined modulo $2\pi$, and we ensure that the shift in $x$ position $\Delta x=\theta(k_y)/2\pi$ satisfies $0\leq \Delta x < 1$. The Wannier states can thus be brought into an order of increasing centre of mass position $\langle \hat X^{cg}\rangle$ by using a single linearized momentum index $J$ \cite{Qi:2011p1804} relating to the parameters of the Wannier state by $K_y = k_y + 2\pi x \equiv 2 \pi J / L_2$, with $J=0,\ldots,N_\text{cell}-1$.

To describe the fractional quantum Hall problem of particles in the lowest Landau-level of a magnetic field, we adopt the Landau gauge $\mathbf{A}=-x B \mathbf{e}_y$, such that our oblique simulation cell is pierced by $N_\phi=N_\text{cell}$ flux quanta, with $L_1 \mathbf{v}_1 \times L_2 \mathbf{v}_2 = 2\pi\ell_0^2 N_\phi$. The periodic Landau-level orbitals $\phi_j(x,y)$, $j=0,\ldots,N_\phi-1$ \cite{Yoshioka:1984p2705,Haldane:1985p2786} are chosen with definite momenta $k_y=2\pi j/L_2$, and achieve their maximum amplitude at $\langle x\rangle = k_y \ell_0^2$.

For the remainder of this manuscript, we choose a specific flat band model to perform a quantitative assessment of the Wannier representation: we work with the Haldane model \cite{Haldane:1988p634}, defined on the lattice shown in Fig.~\ref{fig:HaldaneModel}(a), and choose parameters yielding a nearly flat $C=1$ band: $t_1 = 1$, $t_2 = 0.60$, $t_3 = -0.58$ and $\phi= 0.4 \pi$ \cite{Wang:2011p2572}. 
The resulting Berry curvature, shown in Fig.~\ref{fig:HaldaneModel}(b), is inhomogeneous. 
The definition 
(\ref{eq:Position}) implies that
\begin{equation} 
\frac{\partial}{\partial k_y} \langle \hat X^{cg}  \rangle |_{x} = -\frac{1}{2\pi}\frac{\partial \theta(k_y)}{\partial k_y} = \int_0^{2\pi} \mathcal{B}(p_x,k_y) dp_x,
\label{Bdens}
\end{equation}
i.e., the $k_y$ dependency of the integrated Berry curvature translates into a non-linear evolution of the centre of mass position for the Wannier states, shown for the Haldane model in Fig.~\ref{fig:HaldaneModel}(c). By contrast, the lowest Landau-level has a constant Berry curvature, yielding linear behaviour.

Having defined a single particle basis $\{\phi_j\}$ characterized by a single linear (linearized) momentum index $j$ ($J$) for the lowest Landau-level (FCI), respectively, we can compare the structure of their interaction Hamiltonians by evaluating matrix elements. Formally, two-body interactions can be written in the generic form $\mathcal{H}=\sum_{\{j_i\}} V_{j_1j_2j_3j_4} \hat c^\dagger_{j_1} \hat c^\dagger_{j_2} \hat c_{j_3} \hat c_{j_4}$, with matrix elements $V_{j_1j_2j_3j_4}$ given by the projection to the lowest band \cite{Yoshioka:1984p2705,Regnault:2011p2571}.

\begin{figure}
\begin{center}
\includegraphics[width=0.9\columnwidth]{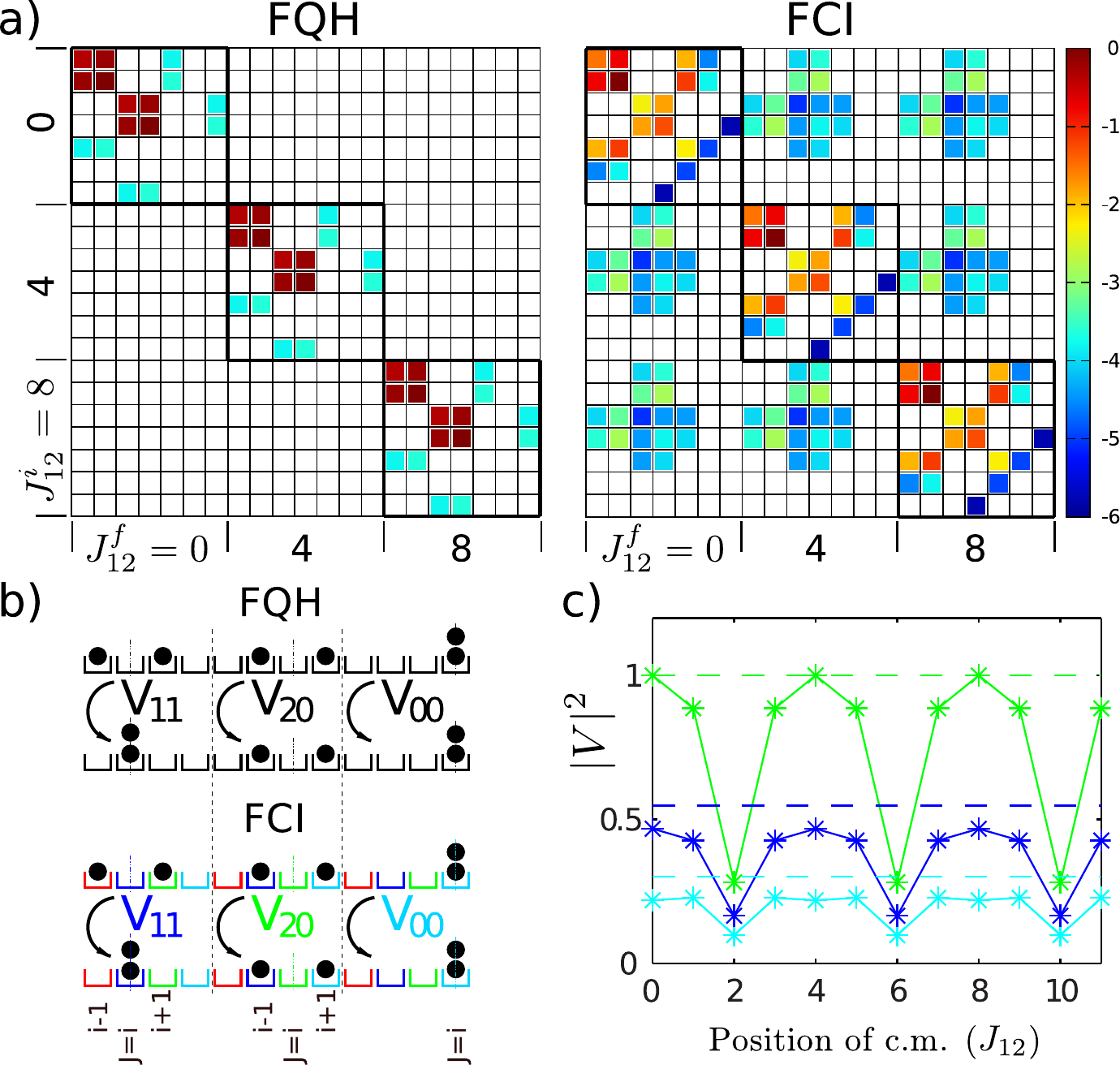}
\caption{\label{fig:MatrixElements}
(a) Magnitudes of matrix elements for the two-body delta interaction between pairs of particles with incoming (outgoing) momenta $J^{i(f)}_{12}=[(J_1+J_2)\!\!\mod N_\text{cell}]$ in a finite size geometry for the FQHE on the torus with $N_\phi=12$ (left) and for the FCI in the lowest band of the Haldane model for a $3\times 4$ lattice (right). (b) Schematic showing some short range interactions, including a `squeezing' process $V_{11}$ as well as two diagonal interaction terms $V_{20}$, $V_{00}$. Panel (c) shows how the magnitude of these processes depends on the centre of mass position for the FCI (solid) as compared to FQH (dashed).}
\end{center}
\end{figure}

We focus on contact interactions for bosons, as this case has a straighforward interpretation for the continuum and on lattices. %
To treat the FCI case, we flatten the residual dispersion of the topological band (ensuring that the Wannier states are energy eigenstates).
Upon comparison, we find that the matrix elements for the FCI Hamiltonian differ from the FQH case in two aspects:
The first issue concerns momentum conservation. For the FQHE, the momentum of the 
Landau-gauge $k_y$ is conserved in scattering processes, and hence $V^\text{FQH}_{j_1j_2j_3j_4}\propto \delta_{j_1+j_2,j_3+j_4}$. In a topological flat band, momenta $q_1$ and $q_2$ are conserved separately while the linearized 
momentum index of the Wannier states $J$ is conserved only modulo $L_2$, i.e., $V^\text{FCI}_{J_1J_2J_3J_4}\propto \delta^{\mod L_2}_{J_1+J_2,J_3+J_4}$. In Fig.~\ref{fig:MatrixElements}(a), we illustrate the magnitude of matrix elements for a small system. The figure clearly shows the block-diagonal structure for the FQH case, reflecting full momentum conservation, while the  FCI Hamiltonian has several off-block-diagonal entries. Nevertheless, both matrices are similar in that the entries of largest magnitude are located at the same positions.
The second difference lies in the translational invariance of the matrix elements. For the FQHE, the amplitude for scattering processes is invariant under translations in momentum space (or effectively in real-space, given that $\langle x \rangle \propto k_y$). For the FCI on the other hand, the non-linear depencency of $\langle \hat X^{cg}(K_y) \rangle$ imprints a variation of the matrix elements with periodicity $L_2$. This effect is illustrated for several nearest neighbor interactions in Fig.~\ref{fig:MatrixElements}(b,c).
Given these two qualitative differences -- momentum conservation and translational invariance of the matrix elements -- the FCI Hamiltonian in the Wannier basis cannot have eigenstates that are identical to those the corresponding FQH problem, as had been conjectured in Ref.~\cite{Qi:2011p1804}.

As a next step, we evaluate the similarity of the wavefunctions for the FCI and FQH problems in terms of their overlap when written in the Wannier and Landau-gauge bases, respectively. We analyze the case of a half filled band, or $\nu=1/2$, for systems with $N=6$, $8$, and $10$ bosons, on lattices of several aspect ratios. For the corresponding FQH problem, we choose a simulation cell with the same geometric features, namely a torus with an aspect ratio given by $R=L_2/L_1$ and opening angle $\gamma=\pi/3$ to match the hexagonal lattice underlying the Haldane model. The FCI Hamiltonian in the Wannier basis is diagonalized in the Fock spaces for total linearized momenta $J_\text{tot}=[(\sum_{n=1}^N J_n) \!\!\mod L_2]$. The Hilbert space for the FQHE has full translational symmetry and segments into blocks with total momentum $j^T_\text{tot}=0,\ldots,N_\phi-1$, (where $N_\phi=N_\text{cell}$). Accordingly, each FCI eigenstate in sector $J_\text{tot}$ can have overlap with several sectors $j^T_\text{tot}$ of the FQH problem satisfying $[j^T_\text{tot}\!\!\mod L_2] = J_\text{tot}$.
In addition, the Laughlin state \cite{Laughlin:1983p301} which is the exact groundstate of contact interactions in the lowest Landau level at filling $\nu=1/2$ has a twofold topological degeneracy $d_\text{GS}=2$. Hence, we calculate the total groundstate overlap $\mathcal{O}$ as an average for both groundstates $|\Psi^\text{FCI}_s\rangle$, taking into account projections $\mathcal{P}_{j^T}$ onto sectors with torus groundstates $|\Psi^T_{s'}\rangle$, yielding
$\mathcal{O}=\frac{1}{d_\text{GS}} \sum_{s,s'=1}^{d_\text{GS}} |\langle \Psi^T_{s'} | \mathcal{P}_{j^T(s')} | \Psi^\text{FCI}_s \rangle |^2$.
For our largest system, $N=10$ particles on a $L_1\times L_2=5\times 4$ lattice with a Hilbert space of $d\simeq 5\times 10^6$, we find a clear gap above two low lying groundstates that yield a total overlap of $\mathcal{O}=0.822$. This value corresponds to Qi's gauge-choice of the Wannier states \cite{Barkeshli:2011p2526}. In addition, we have run numerical optimizations of the phases $\chi(k_y)$ maximizing $\mathcal{O}$, and have found minor changes $\lesssim 1\%$ in the overall result. Thus, we report overlaps conforming with the initial gauge choice \footnote{An analytic gauge fixing procedure was recently proposed \cite{Wu:2012p3204}}.
The FCI states have a total weight $\mathcal{W}=0.885$ within the momentum sectors of torus groundstates, establishing an upper bound for the overlap. The `leakage' of weight outside the groundstate sectors results from off-block diagonal entries in the FCI Hamiltonian, and is independent of the gauge choice.

\begin{figure}
\begin{center}
\includegraphics[width=0.99\columnwidth]{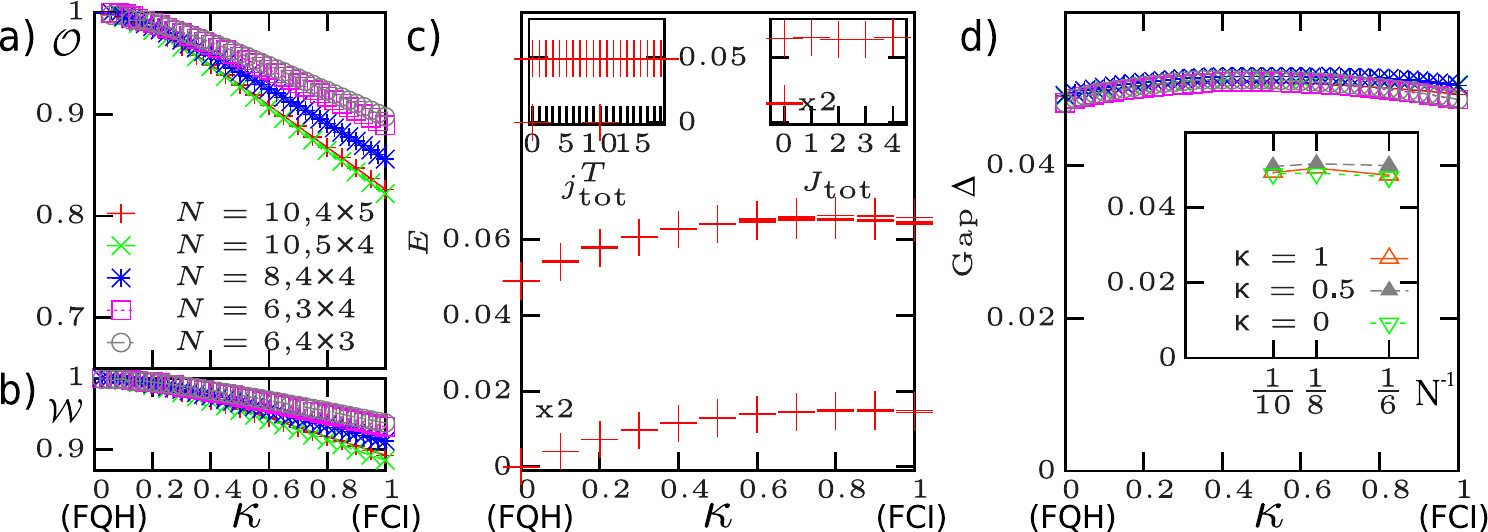}
\caption{\label{fig:AdiabaticSpectrum}(a) Overlap of the groundstate manifolds of $\mathcal{H}(\kappa)$ and $\mathcal{H}^\text{FQH}$ (see text) for bosons at $\nu=1/2$. (b) Total weight in torus groundstate subspace. (c) Spectrum for a system with $N=10$ particles, along a path adiabatically connecting a continuum problem on the torus to the FCI Haldane model on a lattice of $4\times 5$ unit cells. The insets show the momentum-resolved spectrum for the torus (left) and the pure FCI system (right). (d) Gap for several systems of different sizes and aspect ratios. Inset: finite size scaling of the gap.}
\end{center}
\end{figure}

For the Haldane model, we conclude that the Wannier construction yields non-trivial overlaps with the eigenstates on the torus. However, in light of our results, Qi's construction does not yield satisfactorily accurate trial wavefunctions. %
Nevertheless, the Wannier construction allows us to formulate the FCI and FQH problems in a Hilbert space of the same structure, making it convenient not only to calculate overlaps but also to construct an adiabatic continuation between them. Hence, we can formulate a superposition of both Hamiltonians and analyze its spectrum at any intermediate value of an interpolation parameter $\kappa\,$:
\begin{equation}
\label{eq:AdiabaticCont}
\mathcal{H}(\kappa) = \kappa \mathcal{P}_\text{TFB}\left[\mathcal{H}^\text{FCI}(U)\right] + (1-\kappa) \mathcal{H}^\text{FQH}(V_0=1),
\end{equation}
where $\mathcal{P}_\text{TFB}$ denotes the flattening of the topological flat band.
In order to fix the relative energy scales in the two problems, we analyze the magnitude of the gap and choose a value of the onsite interactions $U$ that equalizes its numerical value for $\kappa=0$ and $\kappa=1$, respectively. As we find little scaling of the gap with the system size (see below), we choose a single value of $U=0.2649$ throughout our study. With the definition (\ref{eq:AdiabaticCont}) of the adiabatic continuation, we can analyze the overlap and leakage of the groundstate wavefunctions as a function of $\kappa$. Fig.~\ref{fig:AdiabaticSpectrum}(a) summarizes our results for several system sizes, showing how the overlap drops with increasing system size.

We now examine whether the Laughlin state on the torus is adiabatically connected to the groundstate of the Haldane model with delta interactions. The properties of topologically ordered phases are conserved under the variation of system parameters as long as the groundstate manifold is protected by a finite gap $\Delta$. Hence, we evaluate the spectrum of the class of Hamiltonians (\ref{eq:AdiabaticCont}) as a function of $\kappa$, as displayed in Fig.~\ref{fig:AdiabaticSpectrum}(c) for $N=10$ particles. The spectrum clearly shows a twofold degenerate groundstate, which is well separated by a gap from a continuum of excited states at higher energy. 
To survey finite size scaling, we report $\Delta$ for different lattice geometries in Fig.~\ref{fig:AdiabaticSpectrum}(d). The magnitude of $\Delta$ is found to be weakly dependent on the interpolation parameter $\kappa$. Furthermore, it also depends weakly on system size. The inset shows the finite size scaling of $\Delta$ for different adiabatic continuation parameters, clearly revealing that the gap remains open in the thermodynamic limit. Hence we confirm that the bosonic Laughlin state at $\nu=1/2$ is adiabatically connected to the ground state of the Haldane model, which firmly establishes that they are in the same universality class. We underline that finding a path of adiabatic continuity is a non-trivial task. In our formulation, the choice of the Wannier basis yields the definition for a successful path of deformation. 

\begin{figure}
\begin{center}
\includegraphics[width=0.98\columnwidth]{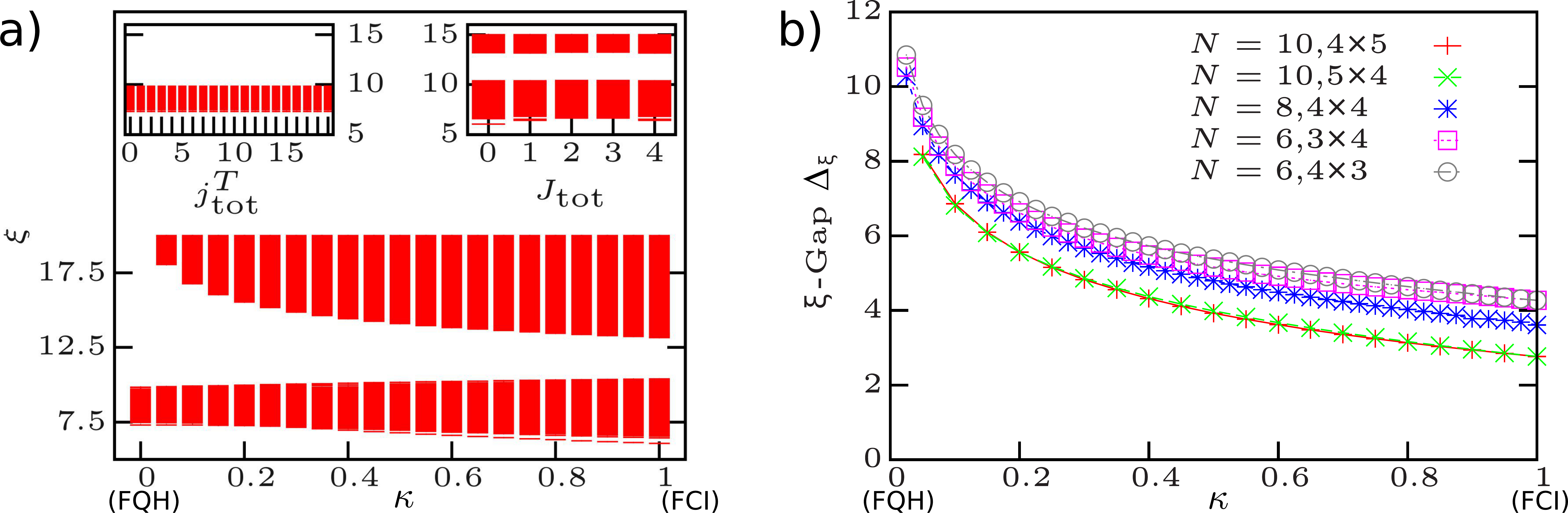}
\caption{\label{fig:AdiabaticPES}
Entanglement spectrum (a) and entanglement gap (b) for the reduced density matrix of a block with $N_A=N/2$ particles. System sizes in analogy with the energy spectrum shown in Fig.~\ref{fig:AdiabaticSpectrum}(c,d).}
\end{center}
\end{figure}

Finally, given its importance for identifying incompressible states in FCI models \cite{Regnault:2011p2571,Wu:2012p2552},
we consider the entanglement spectrum of the ground states along the trajectory $0\leq \kappa \leq 1$. We evaluate the particle entanglement spectrum that encodes the number of quasihole excitations above the groundstate \cite{Zozulya:2007p3279,Sterdyniak:2011p2080}. 
As shown in Fig.~\ref{fig:AdiabaticPES}(a), we find a clear entanglement gap $\Delta_\xi$, and that the count of entanglement eigenstates below the gap remains conserved at the expected universal number \cite{Regnault:2011p2571} within all momentum sectors and at all values of $\kappa$.
The magnitude of $\Delta_\xi$ increases monotonically as the system is deformed from the FCI ($\kappa=1$) towards the FQHE limit ($\kappa=0$), as shown for different system sizes in Fig.~\ref{fig:AdiabaticPES}(b). A quantitative extrapolation of $\Delta_\xi$ to the thermodynamic limit is not justified on our limited data base, but it appears likely that the monotonic behaviour of $\Delta_\xi$ carries over to the thermodynamic limit. %
Hence, our data are consistent with an extended adiabatic continuity in terms of the entanglement gap.

In conclusion, we have established an approach to show that fractional Chern insulators are adiabatically connected to fractional quantum Hall states.  Our technique uses Qi's construction of hybrid Wannier orbitals, and extends to the thermodynamic limit by a robust extrapolation procedure. Specifically, we have used this concept to prove that the FCI ground state of bosons in the half filled Chern band of the Haldane model is in the same universality class as the Laughlin wavefunction at $\nu=1/2$.

\begin{acknowledgments}
We thank N.~R.~Cooper, R.~Roy, B.~Béri, Y.-L. Wu, G. Conduit, and especially N.~Regnault for insightful discussions.
T.S. enjoyed the hospitality of Trinity Hall Cambridge and the Cavendish Laboratory. 
G.M. acknowledges support from the Leverhulme Trust under grant ECF-2011-565 and from the Newton Trust of the University of Cambridge.
\end{acknowledgments}

{\it Note added}. Shortly after publishing the first version of this manuscript, a related preprint has appeared \cite{Wu:2012to}.

\bibliography{wannier-fci}

\end{document}